# Comparison of Characteristics and Practices amongst Spreadsheet Users with Different Levels of Experience

Kenneth R. Baker, Stephen G. Powell, Barry Lawson, and Lynn Foster-Johnson

*Barry.Lawson@dartmouth.edu*

**ABSTRACT**

We developed an internet-based questionnaire on spreadsheet use that we administered to a large number of users in several companies and organizations to document how spreadsheets are currently being developed and used in business. In this paper, we discuss the results drawn from of a comparison of responses from individuals with the most experience and expertise with those from individuals with the least. These results describe two views of spreadsheet design and use in organizations, and reflect gaps between these two groups and between these groups and the entire population of nearly 1600 respondents. Moreover, our results indicate that these gaps have multiple dimensions: they reflect not only the context, skill, and practices of individual users but also the policies of large organizations.

**1. INTRODUCTION**

Spreadsheets are common and important in the world of business. They are used for clerical tasks, for modeling and analysis, and for communication. As they have become more widely accepted, they have been employed for increasingly critical business applications. Nevertheless, researchers, auditors, and consultants frequently express the concern that spreadsheet use defies the norms of discipline that can be found in other business activities. They point out that spreadsheet use implies certain risks and costs; and in that light, there is a need for companies to pay greater attention to the way spreadsheets are managed in the organization.

In the Tuck Spreadsheet Engineering Research Project (SERP), we have examined current organizational practice as it relates to the use of spreadsheets, with the ultimate aim of developing a set of "best practices" for creators and users. An early step in this research has been to document how spreadsheets are currently being used. For this purpose, we developed a detailed questionnaire that could be administered on the internet. The questionnaire has been made available to seven different groups representing three corporations, graduates of two business schools, and affiliates of two software firms. In this paper, we discuss results drawn from these seven surveys with special attention to two distinct subgroups drawn from the entire population of those surveyed: Group A - those who classify themselves as normally working on non-critical and small spreadsheets and with limited expertise, and Group B – those who classify themselves as working on spreadsheets that are 'critical' for their organization and as possessing high expertise and normally working on very large spreadsheets (i.e., over 10,000 cells).

The choice of these two groups was made to identify differences in the contexts within which they work as well as in some of the spreadsheet engineering practices used. These results are also compared with the results for the entire population of respondents. This analysis will highlight the variations among users in different contexts and with different levels of expertise and experience.

The survey responses allow us to consider not only the skills of individual users but also the policies of organizations. Analysis of these responses also suggests other lines of inquiry using the data collected and, perhaps, other surveys to be designed.

In the next section, we review survey results that have appeared in the research literature. In Section 3, we describe the SERP questionnaire and the populations to whom the survey was administered. Section 4 considers the contexts and personal characteristics of the two groups analyzed, and Section 5 focuses on the differences in spreadsheet practices between the two groups. Finally, in Section 6, we discuss the implications of the results and describe other uses of the questionnaire that might shed additional light on spreadsheet use.

## 2. LITERATURE REVIEW

Spreadsheets have been around for over 25 years, but there have been few published surveys that provide a broad-based look at spreadsheet practices. Here are the most important surveys that we were able to find in the relevant research literature.

- Pemberton and Robson (2000) surveyed part-time students (who were working full time) at the University of Northumbria Business School. Of the 227 students surveyed, about 30 did not use spreadsheets, so the effective sample size was 197. The average age was 29. About half the sample (48%) used spreadsheets three or more times a week. The software was Excel (94%), Lotus (5%), and QuattroPro (1%). The survey suggested that most spreadsheet use was unsophisticated, perhaps due to limited amounts of training.

- Panko and Halverson (1996) studied spreadsheet risks and provided an appendix referencing studies describing selected corporate and workgroup policies regarding the creation and testing of spreadsheet models. Their findings suggested that "corporate controls tend to be nonexistent or completely ignored."

- Chan and Storey (1996) surveyed members of a Lotus mailing list in 1992, sending out 1000 questionnaires and receiving 256 returns from business analysts in various functional specialties. The respondents were distributed broadly over several industries. The survey described their training and the most frequent types of analyses they did, along with an indication of the frequency with which they used nine prominent features. The main part of the Chan-Storey article describes a model for the (statistical) relationship among analytic tasks performed, spreadsheet proficiency, use of specific spreadsheet features, use of other software packages, and satisfaction with these software packages. The strongest relationship linked spreadsheet proficiency with the performance of specific tasks. In an expanded version of the model, spreadsheet proficiency and the importance of decisions made were found not to be significantly related.

- Hall (1996) surveyed spreadsheet developers in Australia in late 1991. She sent out 268 questionnaires, received 106 returns, but only 82 of those completed the questions on controls (good practices), which were the focus of the study. The respondents answered questions about a specific spreadsheet project of their choice. The survey examined a list of 55 good practices, obtaining an indication of whether the respondent used or should have used each one.  Some of the practices surveyed are common to those surveyed for this paper, e.g., creating a separate area for data input or using cell protection, although, again, the data are roughly 15 years old.

Two features of this literature review are striking.  First, recent data are lacking.  Except for Pemberton and Robson's survey, these efforts are more than ten years old and were administered to users of spreadsheet software that pre-dates Microsoft Excel, which is the dominant spreadsheet program in use today.  Second, the sample sizes are not large.  At a size of 256, the Chan-Storey sample appears to be the largest in the field, and the other samples are somewhat smaller.

A related literature deals with studies (not surveys) of spreadsheet use in practice.  These articles deal with rather small samples, but they address narrowly-focused research questions.  In specialized ways, they contribute to our understanding of spreadsheet practice.

- Croll (2005) described interviews with about 20 auditors, accountants, bankers, insurers, analysts, and the like, showing how spreadsheets play a critical role in London's financial community.  The interviews suggested a classification of spreadsheets used in practice.  Croll concluded that the awareness and control of risk are uneven, with banking, professional services, and private finance being the most aggressive at dealing with the potential for spreadsheet errors.  His findings provide a useful backdrop for the portion of our results that deal with issues of risk.

- Grossman, Mehrotra, and Özlük (2005) conducted field interviews to identify spreadsheets that were vital to the companies that use them. They identified five classes of such spreadsheets: application software, financial risk management tools, executive information systems, business process infrastructure, and complex analytical tools. In each category, they describe one or several spreadsheets in use. In general, they observe a misalignment between the importance of these spreadsheets and the resources devoted to creating and maintaining them.

- McGill and Klobas (2005) studied 159 end users to test hypotheses related to a multifaceted model of the relationship between success and the extent of developer knowledge and user knowledge.  Developer knowledge was found to be closely related to the perceived quality of the application, whereas user knowledge was found to be closely related to the impact of the application on decision making. Their work is also important in the literature because it advocates objective mechanisms to assess levels of spreadsheet knowledge.

- Lawrence and Lee (2004) presented a report to the Financial Services Forum.  They provided a framework for the analysis of project financing and presumed that the accompanying analysis could apply as well to spreadsheets.  In the appendix, the report summarized the experience of Mercer Finance and Risk Consulting, profiling the 30 largest spreadsheet models that they studied during the preceding year.  Their statistical results provide a benchmark for some of our findings.

- Edwards, Finlay and Wilson (2000) developed a set of guidelines for the "do-it-yourself" spreadsheet creators and a set of best practices for verifying spreadsheets and improving logic and data management.  Some of these practices are reflected in the SERP survey used as a basis for this research.

Comparison of Characteristics and Practices Among Spreadsheet Users
With Different Levels of Experience: Baker, Powell, Lawson, & Foster-Johnson

- Kreie, Cronan, Pendley, and Renwick (2000) studied 66 end users, contacted over the internet, to investigate the question of whether the quality of end-user computing applications could be improved by training end users in analysis and design methods. (Their answer: yes.)

Thus, field work in end-user computing has supplemented broad surveys with specialized portraits of spreadsheet use, within the bounds of narrow research questions posed in experiments and interviews. Such efforts complement surveys by exploring the dimensions of organizational psychology that influence the use of spreadsheets.

A third segment of current literature provides guidance on "recommended" spreadsheet practices. Three of these papers, by PriceWaterhouseCoopers (1999), BPM (2004), and Raffensperger (2004) are good examples of detailed recommended practices. Several of their recommendations are reflected in the SERP questionnaire designed and administered for this project.

**2.1    Purpose of This Paper**

This paper analyzes the results of 1597 responses collected from seven surveys conducted by SERP. It compares the responses to several of the survey questions of two distinct sub-populations, categorized by their relative (self-defined) expertise and experiences with spreadsheets. One purpose is to determine if there are important differences in the spreadsheet practices between these two groups. Another purpose is to see if these differences help us to identify good, if not "best", individual and corporate practices.

Our analysis builds on a recent paper by Baker, Powell, Lawson and Foster-Johnson (2006). The authors compared results from 770 graduates of the Tuck School of Business with a benchmark group of 550 respondents representing a self-selected group of people on the mailing lists of two software development companies. One of the results suggested that spreadsheet designers and users with considerable experience employ more and different practices than those with less experience.

**3. THE SERP SURVEY**

Our questionnaire was organized around a seven-stage life cycle model for a typical spreadsheet. The stages in this life cycle are: designing, testing, documenting, using, modifying, sharing, and archiving. (As in most life cycle models, the typical path is not serial. Instead, a spreadsheet may revisit a previous stage—perhaps several times—during its useful life.) The questionnaire contained questions in each of these areas; in addition, it contained questions on training, quality control, and risk as they relate to spreadsheets. Finally, there were questions that described the respondents themselves. In all, the questionnaire contained 67 items, some of which were open-ended, and took about 15-20 minutes to complete.

Seven populations were invited to fill out the questionnaire. The largest population consisted of alumni from the MBA Program at the Tuck School of Business. Invitations were emailed to approximately 7000 alumni, and responses were obtained from 770. A comparable population consisted of alumni from the London Business School and resulted in 76 responses. Three populations represented targeted private corporations with 41, 67 and 75 responses respectively. The per cent response rate in these last four surveys is difficult to determine, but in at last two of the cases, the respondents represented over 50% of the possible number of respondents.

Comparison of Characteristics and Practices Among Spreadsheet Users
With Different Levels of Experience: Baker, Powell, Lawson, & Foster-Johnson

The other two populations were people whose names appeared on mailing lists generously provided by two software companies, Frontline Systems, Inc. and Decisioneering, Inc. Frontline produces versions of the Excel Solver for optimization of spreadsheet models, and Decisioneering produces Crystal Ball, an Excel add-in for Monte Carlo simulation. Together, these two supplementary samples contained 509 respondents from approximately 20,000 invitations sent out. We anticipated that these two samples would represent a relatively sophisticated and technical set of spreadsheet users because of their association with optimization and simulation applications.

As mentioned earlier, our survey contained 67 questions. We limit our discussion to those questions that shed light on the context and characteristics of the respondents and their spreadsheet design and use practices. For a detailed summary of results, interested readers can visit our project's website: (http://mba.tuck.dartmouth.edu/spreadsheet/index.html).

**3.1 Defining Two Groups to Study**

While we designed the questionnaire based on the life cycle in the creation and use of spreadsheets, the main focus of this paper is on respondent characteristics and practices. To accomplish our purposes it was appropriate to identify within the total number of respondents to the SERP survey two subgroups: one comprised of those who had the least experience and expertise, the other representing those with the most experience and expertise. We also assumed that the size of spreadsheets that respondents normally designed and used would further reflect differences in skills, spreadsheet tools, and practices. Our hypothesis was straightforward: Those respondents who were most experienced, had the highest level of expertise, and normally worked on larger, more complex and important spreadsheet models, would use the best practices. Our goal was to identify these practices.

The findings have been separated into responses from Group A and Group B and often compared with the responses from all survey respondents. We have focused on responses to those questions that reflect the differences between the two groups with regard to selected characteristics and their spreadsheet practices. Therefore, in order to select two groups for comparison, we used responses to three of the survey questions to define each group.

> **What level of importance do spreadsheets have in your job?**
> Group A - a. Unimportant; b. Moderately important; c. Very important;
> Group B - d. Critical
>
> **Please classify your experience with spreadsheets.**
> Group A - a. Little or no experience; b.. Some experience; still a beginner;
> c. Extensive experience; some expertise;
> Group B - d. Very experienced; high expertise.
>
> **How large are the models you normally create?**
> Group A - a. under 100 cells; b. 101 to 1000 cells; c. 1001 to 10,000 cells;
> Group B - d. 10,001 to 100,000 cells; e. over 100,000 cells

Group A consisted of those who said:
    a. the level of importance spreadsheets have in their job is either "unimportant",
        "moderately important" or "very important" AND

Comparison of Characteristics and Practices Among Spreadsheet Users
With Different Levels of Experience: Baker, Powell, Lawson, & Foster-Johnson

    b. their experience with spreadsheets was "little or no experience", "some experience; still a beginner" or "extensive experience; some expertise" AND
    c. the sizes of models normally created are under 10,000 cells.

Group B consisted of those who said:
    a. the level of importance spreadsheets have in their job is "critical" AND
    b. their experience with spreadsheets could be characterized as "very experienced; high expertise" AND
    c. the sizes of models normally created normally exceed 10,000 cells.

One would guess that spreadsheets are important to most of those who responded to the survey. In fact, we asked about the "level of importance for spreadsheets in your job," and less than 1% replied that spreadsheets were "unimportant". One might infer that there is only a subjective difference between someone with "extensive experience – some expertise" and another person who is "very experienced – high expertise". However, this difference, when combined with the other two characteristics (i.e., relative importance and size of the spreadsheets on which they worked), proved to distinguish clearly the two groups. Of the 1597 total respondents, 165 (10.3%) were in Group A, 175 (10.9%) in Group B. "All" responses are those from the total of 1597 respondents.

The figure below shows their distribution among the seven participating survey audiences. We make no claim that the respondents to the survey (including Group A and Group B) represent typical spreadsheet designers and users. Most come from the business or corporate world and have advanced degrees and considerable experience. We can assume that the gap between the typical designer and user may be greater than that portrayed in this analysis. How great this gap is remains a subject for later inquiry.

Figure 1. Percentage of each of seven survey audiences represented in Groups A and B.

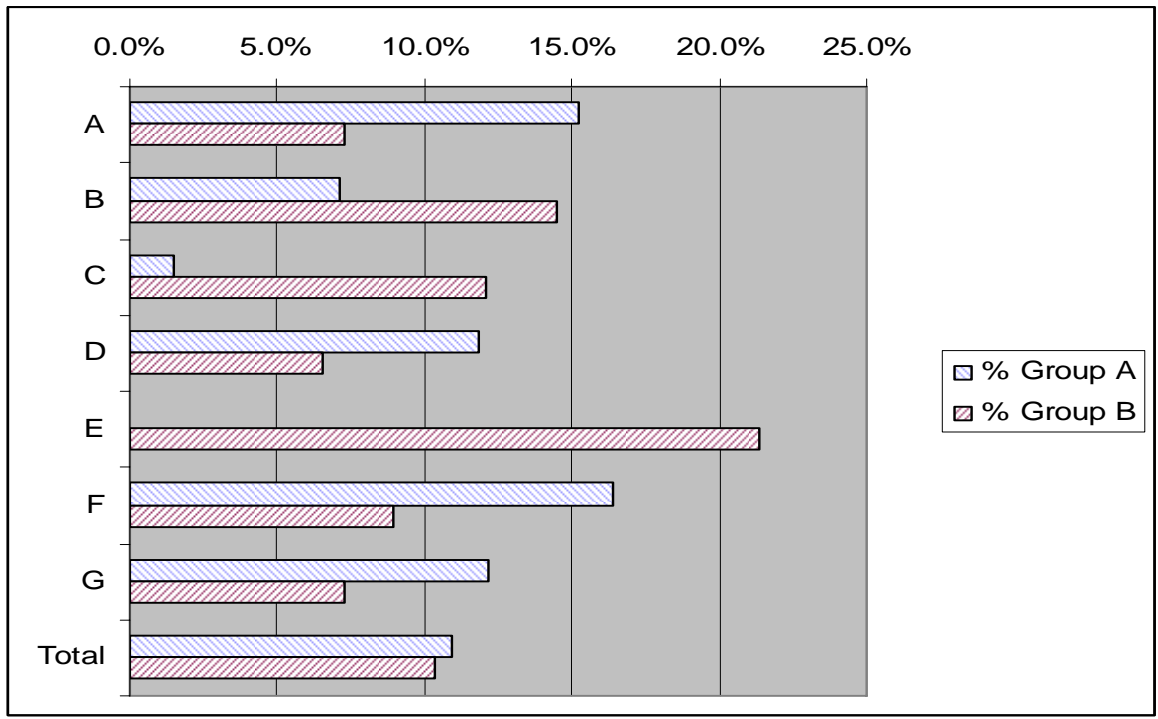

Comparison of Characteristics and Practices Among Spreadsheet Users
With Different Levels of Experience: Baker, Powell, Lawson, & Foster-Johnson

At least five percent of each survey audience is represented in Group B, with the largest (Survey E) exceeding 20%. All but one (again Survey E) is represented in Group A. Four audiences have larger percentages of Group A than Group B representatives.
The average age of our respondents was about 40, indicating that the respondents had typically been out of school for several years. Over three-quarters characterized their position as being either "supervisor or manager" or "executive." About 53% worked in service industries, with another 18% in manufacturing.

## 4. CHARACTERISTICS OF GROUP A AND GROUP B

In order to characterize the members of the groups we selected several questions, in addition to the three used to define the two groups.

**Time Spent With Spreadsheets**

Nearly ninety-three percent of Group A spends less than 25% of their time on spreadsheets while over eighty-eight percent of Group B spend more than 25% of their time on spreadsheets (and over twenty-six percent devote more than 75% of their time). Hence there is a wide gap in the commitment of time and, hence, the experience gained from this commitment.

Table 1. Time Normally Spent on Spreadsheets Per Week, by Group.

|         | Group A | Group B | All   |
|---------|---------|---------|-------|
| 0-25%   | 92.6%   | 11.5%   | 44.7% |
| 26-50%  | 6.9%    | 24.2%   | 30.4% |
| 51-75%  | 0.6%    | 37.6%   | 17.8% |
| 76-100% | 0.0%    | 26.7%   | 7.2%  |

**Number of Spreadsheets Used in a Week**

Similarly Group B respondents use many more spreadsheets in a normal week. While 85.1% of Group A uses five or fewer spreadsheets, 78.2% of Group B uses more than five (and 61.8% use more than 10 spreadsheets). Table 2 confirms the intensity of spreadsheet use in Group B as compared to Group A.

Table 2. Number of Spreadsheets Normally Used Per Week, by Group.

|             | Group A | Group B | All   |
|-------------|---------|---------|-------|
| 0-1         | 21.1%   | 0.0%    | 5.8%  |
| 2-5         | 64.0%   | 21.8%   | 40.2% |
| 6-10        | 13.1%   | 16.4%   | 25.6% |
| more than 10| 1.7%    | 61.8%   | 28.3% |

**Main Purposes of Spreadsheets**

Figure 2 portrays the differences between these two groups in the uses for their spreadsheets. Analyzing data is the major purpose for both groups, but Group B respondents indicate that analyzing data, determining trends, and evaluating alternatives are main purposes much more frequently than those in Group A.

Figure 2. Main Purposes of Spreadsheets Used, by Group.

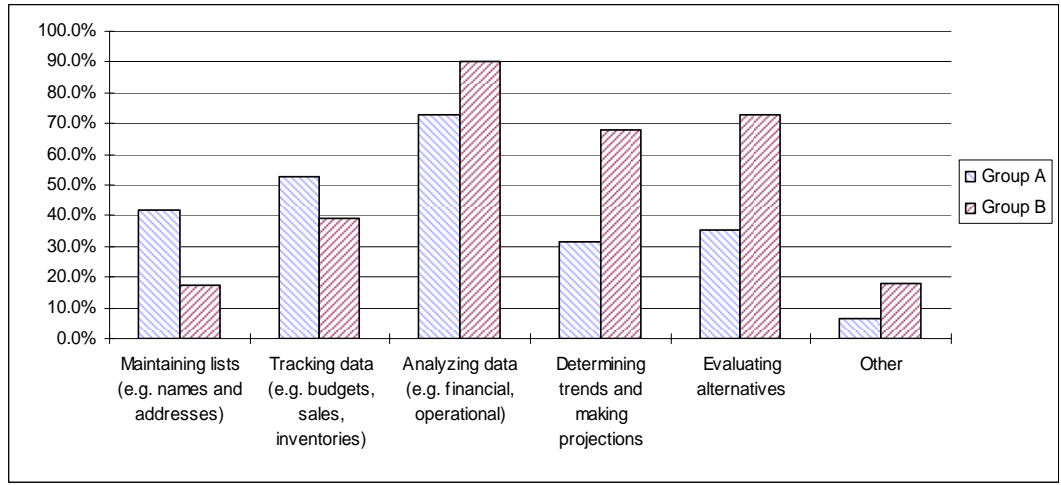

**Users of Spreadsheets One Creates/Uses**

Group A individuals tend to create spreadsheets for smaller numbers of people than do Group B individuals, as the following table shows. Significantly, responses to another question indicate that 47.4% of Group B (but only 2.3% of Group A respondents, and 15.7% of all respondents) create spreadsheets that often become "permanent assets" for their organization. In still another survey question, 56.2% of Group A respondents indicate that the average lifetime of the models they use is less than a year; conversely 70.4% of Group B indicate an average model lifetime of more than a year.

Table 3. Number of Users of Spreadsheets Created, by Group.

|  | Group A | Group B | All |
|---|---|---|---|
| None | 23.4% | 8.2% | 13.5% |
| 1 other person | 29.1% | 10.1% | 18.8% |
| 2-5 other people | 41.1% | 52.8% | 49.9% |
| More than 5 other people | 6.3% | 28.9% | 17.7% |

**Risks and Risk-Avoidance Strategies**

There are reasons why Group B respondents are as committed to spreadsheet design and use as is suggested by survey results. 74.0% report that there is a medium to high risk posed by spreadsheets in their organizations compared to only 30.4% for Group A (and 54.9% for all respondents). Awareness of risk, however, in both groups is less than one might expect. While 84.9% of Group B report "some" or "full" awareness, only 29.6% indicates "full" awareness. For Group A the comparable data are 55.6% and 12.9% for all respondents. Only 16.6% of Group A indicates that their organizations are fully aware of the risks posed by spreadsheets.

**Functional Area of Job**

The functional area within organizations is likely to influence the practices used by respondents. For example, over forty-six percent of Group B respondents indicate that finance was their functional area, while only a little under eleven per cent fit into that category in Group A. The type of spreadsheet functions used, the commitment to spreadsheet accuracy and quality, awareness of risk, and other factors could be largely influenced by this relative domination by the financial service workers.

Table 4. Distribution of Respondents, by Function, by Group.

| Function | Group A | Group B | All |
|---|---|---|---|
| Sales and Distribution | 8.4% | 2.5% | 4.4% |
| Marketing | 18.6% | 3.8% | 10.9% |
| Operations/Manufacturing | 10.2% | 9.6% | 9.5% |
| Engineering and Research | 13.2% | 18.6% | 19.8% |
| Finance | 10.8% | 46.8% | 30.2% |
| Human Resources | 4.2% | 1.3% | 1.3% |
| Other | 34.7% | 17.3% | 23.9% |

**Demographic Characteristics**

Finally, in characterizing the two groups we used two parameters: gender and age. Group A tends to be more heavily populated by females and slightly older than those in Group B. 56.9% of Group A is over age 40; 56.1% of Group B is age 40 or younger.

Table 5. Gender and Age Characteristics of Group A and Group B

| Gender | Group A | Group B | All |
|---|---|---|---|
| Male | 70.1% | 91.4% | 83.3% |
| Female | 29.9% | 8.6% | 16.7% |

| Age | Group A | Group B | All |
|---|---|---|---|
| 20-30 | 15.5% | 16.0% | 13.7% |
| 31-40 | 27.6% | 40.1% | 38.5% |
| 41-50 | 26.4% | 30.9% | 26.2% |
| 51-60 | 20.7% | 9.3% | 14.7% |
| Over 60 | 9.8% | 3.7% | 6.9% |

**Summary of Characteristics**

Group A and Group B differ considerably in several characteristics. Group B individuals spend more time working on more spreadsheets, work more on spreadsheets designed for serve analytical and evaluative functions, have more users for their spreadsheets, and tend to collaborate with more people. Moreover, the individuals in Group B also tend to be younger and more highly represented by those serving financial functions where, one might assume, there may significant pressure to be skilled, accurate, and to use more sophisticated practices. The following section considers various types of practices used by individuals in these two groups.

**5. SPREADSHEET PRACTICES**

The next focus for analysis is to determine if there are differences between these two groups regarding their spreadsheet design and use practices. Several questions included in the online survey relate to such practices as the type of training sought and received; and how spreadsheets are created, tested, documented and shared.

**Training and Work Style**

While there were some differences between the two groups, in general training was a soft spot for both groups and for all respondents and their organizations. Training programs are an exception rather than the rule, with no more than a few days of training each year generally offered in most organizations. The most often repeated reason for lack of training was "lack of time."

When asked what types of training respondents have had in their careers, there were some subtle differences between Group A and Group B (as well as for all respondents). As the following table shows, Group B individuals have received more "occasional informal training," more "demonstrations from colleagues," and significantly more training from "books and manuals" than Group A. This informal and self-taught learning and reference system seems to characterize the spreadsheet training of Group B.

Table 6. Types of Training Reported by Respondents, by Group.

|  | Group A | Group B | All |
|---|---|---|---|
| Books and manuals | 44.6% | 73.3% | 53.6% |
| Demonstrations from colleagues | 52.0% | 58.2% | 52.3% |
| Formal classroom instruction | 41.7% | 40.6% | 37.7% |
| Occasional informal training sessions | 29.1% | 34.5% | 29.2% |
| None | 21.1% | 12.7% | 17.6% |

With regard to work style, both groups overwhelmingly tend to work independently, but there is a decided subset of Group B who works "in a team of 2 or 3." 7.3% of Group B respondents work in larger teams, but only a small fraction of Group A individuals does. In response to another survey question, 28.1% of Group B says they work with either a peer group or a project team in creating spreadsheets. This contrasts with only 4.0% of Group A respondents. 14.8% of all respondents say they work in these peer groups or project teams. The fact remains that most individuals in both groups work independently.

Table 7. Work Styles of Respondents, by Group.

|  | Group A | Group B | All |
|---|---|---|---|
| By yourself | 90.3% | 68.3% | 81.1% |
| In a team of 2 or 3 | 8.0% | 24.4% | 16.3% |
| In a larger team (4 or more) | 1.7% | 7.3% | 2.6% |

**Spreadsheet Design**

Comparison of Characteristics and Practices Among Spreadsheet Users
With Different Levels of Experience: Baker, Powell, Lawson, & Foster-Johnson

Several survey questions addressed how spreadsheets are created by respondents. The responses to three of these questions are presented in Table 8 below.  These concern how often spreadsheets are created "from scratch," how often spreadsheet models are divided into separate, integrated modules, and how frequently data inputs are separated from formulas in spreadsheets.  The responses to these questions collectively demonstrate that Group B individuals more often create spreadsheets from scratch and use good design practices as well.

Table 8.  Selected Spreadsheet Design Practices, by Group

| Do you create spreadsheets from scratch? | Group A | Group B | All |
|---:|---|---|---|
| Always | 37.1% | 48.8% | 36.3% |
| Sometimes | 60.6% | 49.4% | 62.1% |
| Never | 2.3% | 1.8% | 1.5% |

| Do you divide your spreadsheets into integrated modules? | Group A | Group B | All |
|---:|---|---|---|
| Always | 2.3% | 51.8% | 20.4% |
| Usually | 32.4% | 35.4% | 42.6% |
| Sometimes | 51.4% | 12.2% | 32.7% |
| Never | 13.9% | 0.6% | 4.2% |

| Do you separate data inputs form formulas in spreadsheet? | Group A | Group B | All |
|---:|---|---|---|
| Always | 8.1% | 43.6% | 22.3% |
| Usually | 32.4% | 40.0% | 41.4% |
| Sometimes | 48.0% | 14.5% | 31.1% |
| Never | 11.6% | 1.8% | 5.2% |

The table below shows the typical first step in creating a spreadsheet.  The first steps do not vary much among the three groups, although there is less likelihood that Group B individuals would start by "directly entering data into the computer."  Even for this group, however, it is more often the practice than any of the other options offered.

Table 9.  First Step in Creating Spreadsheets, by Group

| | Group A | Group B | All |
|---:|---|---|---|
| Enter the data and formulas directly into a computer | 54.9% | 37.5% | 48.7% |
| Borrow a design from another spreadsheet | 23.4% | 25.0% | 22.8% |
| Sketch the spreadsheet on paper | 14.9% | 20.6% | 17.4% |
| Write the fundamental relationships using algebra | 2.9% | 8.8% | 5.8% |
| Other | 4.0% | 8.1% | 5.3% |

**Use of Software Features**

One of the distinguishing factors between Group A and Group B respondents is the use of Excel features (e.g., functions and tools).  A more extensive working knowledge of a variety of features available through Excel can enhance the sophistication and creativity of the designer and user.

Comparison of Characteristics and Practices Among Spreadsheet Users
With Different Levels of Experience: Baker, Powell, Lawson, & Foster-Johnson

Those who are involved in larger, more complex and critical spreadsheets will require a larger toolkit to create their models and fulfill their models' requirements.

The survey sought information on the relative frequency with which respondents use each of fourteen Excel-related features. Respondents were asked to indicate the frequency of use in terms of the following options: never use, rare use, occasional use, frequent use and daily use. Assigning weights of one to five, respectively, for each of these options, we created a weighted or average frequency of between one and five for each feature – for both Group A and B respondents as well as for all respondents.

Figure 3 presents the result of this analysis, in order from most to least frequent based on the responses of All respondents.

Figure 3. Relative Frequency of Use of Selected Excel Features, by Group

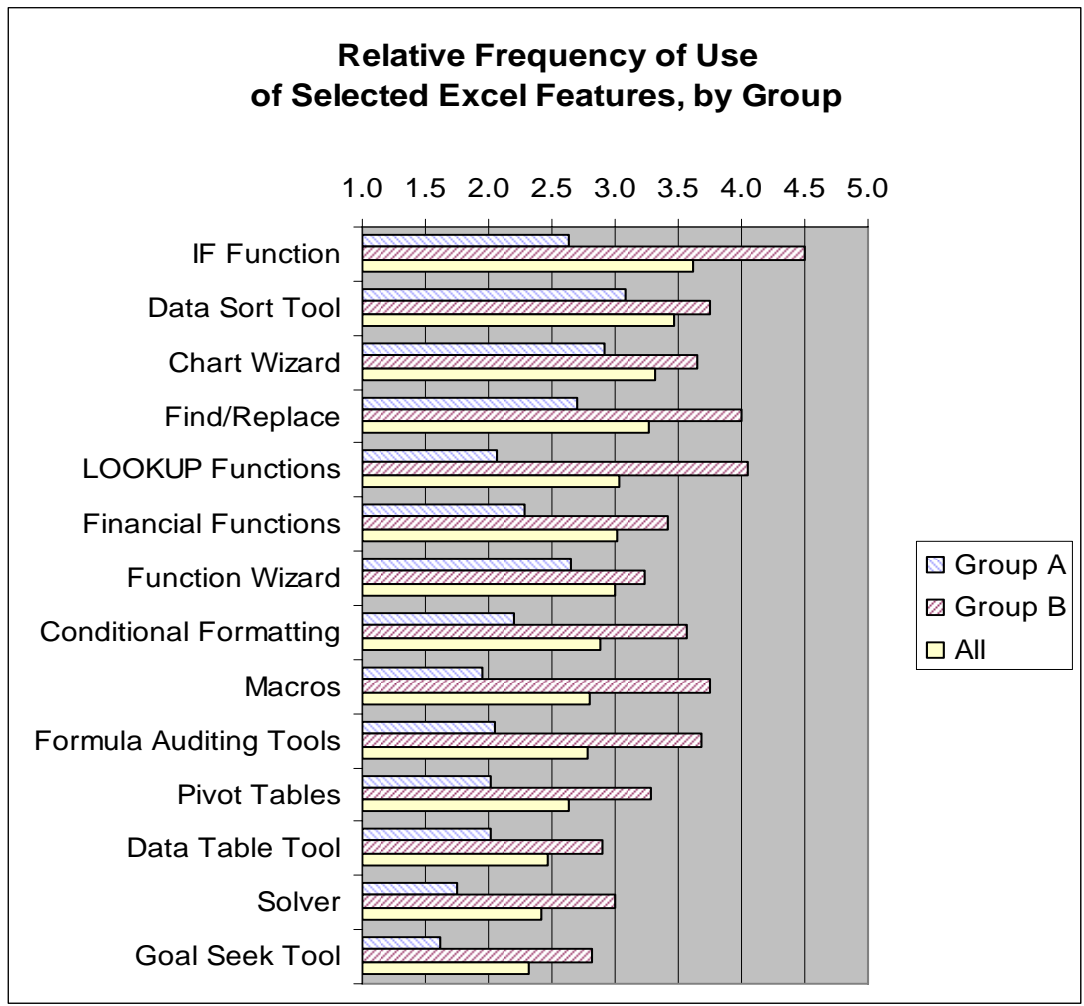

Not only are all of these Excel features used by Group B more frequently than they are by Group A, all but two of the tools are used by Group B individuals, on average, at least at the "occasional use" level, represented by 3.0 on the scale in Figure 3. The Data Sort Tool is the only feature is used by Group A respondents, on average, at that level.  A clear defining measure of the difference in the spreadsheet engineering practices of the two groups of spreadsheet users surveyed is the relative use of these features.

**Evaluating Spreadsheets**

The table below shows considerable differences in the manner by which Group A and Group B approach reviewing, evaluating and/or testing their spreadsheets.  Over 50% of Group B respondents always test their models while only 8% of Group A respondents does.  This undoubtedly reflects to some degree the relative size, complexity and importance of Group B's models.

Table 10.  Frequency of Testing Models Created or Used, by Group

|  | Group A | Group B | All |
|---|---|---|---|
| Always | 8.0% | 53.9% | 24.2% |
| Usually | 18.3% | 25.5% | 26.7% |
| Sometimes | 33.1% | 18.8% | 31.9% |
| Never | 40.6% | 1.8% | 17.1% |

A second way to test could involve using commercial auditing software increasingly available in the marketplace.  Somewhat enlightening is the fact that no one in Group A is in an organization that utilizes audit software, and only 7.1% of those in Group B are.  A third way to evaluate models, and the more traditional approach, is to use a range of techniques as shown below.  Group B takes more advantage of <u>all</u> these techniques than those in Group A or all respondents.  Moreover, the average Group B respondent also uses four of these approaches while the average Group A respondent uses only two.  Again, this finding reflects the size, complexity and relative importance of spreadsheet models created and used by Group B respondents.

Table 11.  Types of Model Evaluation Used by Respondents, by Group.

|  | Group A | Group B | All |
|---|---|---|---|
| Use common sense | 45.7% | 80.6% | 67.4% |
| Test extreme case | 23.4% | 67.9% | 45.9% |
| Examine formulas individually | 34.3% | 65.5% | 45.6% |
| Test performance for plausibility | 24.6% | 64.2% | 43.4% |
| Formula Auditing Toolbar | 9.7% | 51.5% | 28.0% |
| Use a calculator to check selected cells | 29.7% | 46.7% | 38.4% |
| Display all formulas | 18.3% | 21.8% | 18.2% |
| Use Go To - Special | 0.6% | 17.0% | 6.3% |
| Error Checking option | 4.6% | 16.4% | 10.2% |
| Other tools: | 4.0% | 18.2% | 7.6% |

**6. IMPLICATIONS OF THE RESEARCH**

The results of this analysis of SERP survey responses underscore the fact that spreadsheet practices vary substantially from person to person and organization to organization.  Some of these differences relate to the relative importance of the spreadsheets to that person or organization, the level of expertise of the designer and user as well as the size and complexity of the spreadsheet itself.  Other differences may reflect the context within which the spreadsheet creator works. This paper has shown that there are measurable differences in the practices of the two groups of designers and users studied here. These include the type of training undertaken, the work styles, specific design and creation practices the types of tools used, and the methods used (and frequency of that use) to test spreadsheets.

The question of what constitutes "best corporate practices" may be a function of the use to which a spreadsheet is put, its size and complexity, the degree to which it is used by other people, and its importance to the designer, user and organization. There may be a few hard and fast rules or practices, but the context of the use of the spreadsheet is significant.

In this paper we have explored one aspect of spreadsheet practices – how they vary among designers and users in considerably different contexts.

We have identified some practices that are being used to improve the quality of spreadsheets, namely, being better trained, working more closely with colleagues or in a team, not being so quick to start the design by just entering data into a computer, separating spreadsheets into integrated modules, separating data from formulas, and employing more testing methods more frequently. We are endeavoring to establish the most significant "best corporate practices" to guide design and use in the future. Many of these are important for spreadsheet designers; others may require establishing organizational standards and procedures for training of personnel, for the design, testing and review (i.e., quality control) of spreadsheets, and eventually for the role and significance of the use of these spreadsheet models in corporate decision-making.

To deal with these issues, the most effective place to start is with management policies. Putting in place a set of procedures to mitigate spreadsheet risk is analogous to implementing a quality management program: standards and training, for example, could represent key parts of the initiative. A deeper awareness of spreadsheet risk would also filter down to the elements of spreadsheet design and use, leading to more disciplined testing and documentation, greater motivation to use protections, and enhanced use of auditing software.

We plan to compare different populations to see whether these responses were typical and if our survey instrument could be used in a variety of other settings. We also intend to continue our analyses of the data already acquired, defining particular subgroups of the total respondents and examining differences in spreadsheet design and uses practices among them.

Blank page